# Performance Evaluation of an OMPR Algorithm for Route Discovery in Noisy MANETs

Hussein Al-Bahadili[1] and Rami Jaradat[2]

[1]The Arab Academy for Banking & Financial Sciences, Jordan
Hbahadili@aabfs.org
[2]The Hashemite University, Jordan
Rami@hu.edu.jo

**ABSTRACT**

*It has been revealed in the literature that pure multipoint relaying (MPR) algorithms demonstrate both simplicity and outstanding performance, as compared to other flooding algorithms in wireless networks. One drawback of pure MPR algorithms is that the selected forwarding set may not represent the optimum selection. In addition, little efforts have been carried-out to investigate the performance of such algorithms in noisy mobile ad hoc networks (MANETs) suffering from high packet-loss and node mobility. In this paper, we develop and evaluate the performance of an optimal MPR (OMPR) algorithm for route discovery in noisy MANETs. The main feature of this new algorithm is that it calculates all possible sets of multipoint relays (MPRs) and then selects the set with minimum number of nodes. The algorithm demonstrates an excellent performance when it is compared with other route discovery algorithms as it achieves the highest cost-effective reachability.*

## 1. INTRODUCTION

A routing protocol is part of the network layer software that is responsible for deciding which output path a packet should be transmitted on. Many routing protocols have been proposed for MANETs [1, 2]. These algorithms differ in the approach they use for searching a new route and/or modifying a known route, when nodes move. Furthermore, each of the available routing algorithms has its own unique characteristic strengths and weaknesses.

Routing protocols can be classified into different categories according to their properties and applications [2]. The most widely used mechanism for categorizing routing protocols is the one that is based on routing information update, which according to it, routing protocols can be classified into three major categories, these are proactive (static), reactive (dynamic) routing protocols, and hybrid routing protocols [1-4].

Reactive protocols consist of two main phases: route discovery and route maintenance. Route discovery is the process that allows a node in a network to dynamically discover a route to other nodes in the network, either directly within the wireless transmission range, or through one or more intermediate nodes [5]. Reactive protocols such as dynamic source routing (DSR) [2], ad hoc on-demand distance vector (AODV) [3], zone routing protocol (ZRP) [4], and location aided routing (LAR) [5], or variations of them are widely used in MANETs.

It is usually assumed that the cost in terms of bandwidth and power consumptions and delay of information exchange during route discovery, when the knowledge of the route is imperfect, is higher than the cost of point-to-point data forwarding after that knowledge has been acquired. Therefore, the process of route discovery should be done with minimum complexity, overhead, and bandwidth and power consumption [6].

85

Route discovery is used when a source node desires to send a packet to some destination node and does not already have a valid route to that destination; in which the source node initiates a route discovery process to locate the destination. It broadcasts a route request (RREQ) packet to its neighbours, which then forward the request to their neighbours, and so on until the expiration of the RREQ packet. During the forwarding process, the intermediate nodes record in their route tables the address of the node from which the first copy of the broadcast packet is received. Once the RREQ reaches the destination, the destination responds with a route reply (RREP) packet back to the source node through the route from which it first received the RREQ [1].

Pure flooding is one of the earliest, simplest, and reliable mechanisms proposed in the literature for route discovery in MANETs [7-9]. In pure flooding, each node rebroadcasts the message to its neighbours upon receiving it for the first time, starting at the source node. Although it is simple and reliable, pure flooding is costly where it costs $n$ transmissions in a network of $n$ reachable nodes. In addition, pure flooding in wireless networks, using the IEEE 802.11 protocol, results in serious redundancy, contention, and collisions in the network; such a scenario has often been referred to as the broadcast storm problem [10].

To eliminate the effects of broadcast storm problem during route discovery in MANETs, a variety of flooding optimization algorithms have been developed, such as multipoint relaying (MPR) [11-16], probabilistic [7-9], counter-based [10], distance-based [10], locations-based [5, 10], and cluster-based [17, 18] algorithms. They all try to limit the number of collisions by limiting the number of retransmissions. As the number of retransmissions required for broadcasting is decreased, the bandwidth is saved and contention and node power consumption are reduced, and this will improve the overall network performance.

In this work we will focus on MPR algorithms for flooding optimization during route discovery in MANETs. The basic idea behind MPR algorithms is to define a set of nodes called MPRs or relay nodes for each node in the network; these relay nodes are a subset of the one-hop neighbors of the node. They are responsible for forwarding the RREQ packet upon receiving it for the first time, while non relay nodes will not forward the message, so that the packet will be propagated to the whole network to maintain the highest possible reachability with less number of retransmissions.

Different heuristics have been developed in the literature to select the set of relay nodes in MPR algorithms. According to the heuristic that is used, MPR algorithms can be categorized as pure MPR algorithms connected dominating set MPR (CDS-MPR) algorithms, or quality-of-service MPR (QoS-MPR) algorithms [11].

It has been revealed in the literature that pure MPR algorithms demonstrate both simplicity and outstanding performance, as compared to other flooding algorithms that are commonly used in wireless networks. One drawback of pure MPR algorithms is that the selected forwarding set may not represent the optimum selection. In addition, little efforts have been carried-out to investigate the performance of such algorithms in noisy MANETs suffering from high packet-loss and node mobility.

In this paper, we develop a new heuristic for selecting the optimal MPRs set to forward RREQ packets during route discovery in noisy MANETs. Therefore, this MPR algorithm is referred to as optimal MPR (OMPR) algorithm. A noisy MANET is characterized by its high packet-loss, which is expressed in terms of probability of reception. Probability of reception is defined as the probability of a packet being successfully received by adjacent nodes. In order to evaluate the performance of the OMPR algorithm, a number of simulations were simulated using the MANET simulator (MANSim) [19, 20]. The simulations were aimed to estimate the variation of number of retransmissions (RET) and network reachability (RCH) with probability of reception. In addition, in this paper, the performance of the OMPR algorithm is compared against other flooding optimization algorithms, such as: pure flooding [7], probabilistic flooding [7, 8, 9], LAR-1 [5], and LAR-1-probabilistic (LAR-1P) [21] algorithms.

The rest of the paper is organized as follows. Related work is discussed in Section 2. The wireless network environments are defined in Section 3. The concept, cost, and the algorithm for selecting the MPRs in pure MPR algorithm are given in Section 4. The proposed heuristic for selecting the optimum MPRs and the algorithm main features are presented in Section 5. Results of the simulations are presented and discussed in Section 6. Finally, in Section 7, conclusions are drawn and recommendations for future work are pointed-out.

## 2. LITERATURE REVIEWS

In this section, we review some of the research activities and development stages that are related to the MPR algorithms. Self and dominant pruning flooding methods was proposed in [22]. An approach to optimize the performance of flooding broadcast in multi-hop ad hoc networks, namely, the lightweight and efficient network-wide broadcast (LENWB) protocol, was proposed in [27].

A proactive routing protocol for MANETs, namely, the optimized link state routing (OLSR) protocol that employs periodic exchange of messages to maintain topology information of the network at each node was developed and evaluated in [15, 16]. OLSR uses the MPR technique to efficiently and economically flood its control messages and it provides optimal routes in terms of number of hops, which are immediately available when needed. The protocol is best suitable for large and dense MANETs. In [14], pure MPR algorithm for flooding broadcast optimization in mobile wireless networks was developed and analyzed, which showed an excellent performance in sparse networks.

An efficient ad hoc broadcast protocol (AHBP) similar to MPR was developed in [24]. In AHBP, all nodes need to know about their one and two-hop neighbors, and only nodes that are selected as a broadcast relay gateway (BRG) within a broadcast packet header are allowed to rebroadcast the packet. The work in [25] showed that MPR can be used as well in reactive protocols in order to save overhead in route discovery. They specified a simple reactive protocol called MPR distance vector (MPRDV) protocol. In MPRDV RREQs and RREPs are all flooded via MPRs.

An algorithm for computing MPRs, the Mini-ID MPR algorithm was described in [13]. The Mini-ID algorithm is far from optimal but it has the advantage that a node can detect by itself whether or not it belongs to the MPR set of a neighbor. It consists of selecting the nodes in the increasing order of their ID's (or any arbitrary increasing order). They also proposed a CDS election algorithm based on MPR called (MPR-CDS). Unlike MPR, MPR-CDS algorithm does not require the last hop knowledge. The proposed algorithm requires a total ordering of the nodes. In this algorithm, a node decides that it is in the CDS if and only if the node is smaller than all its neighbors, or it is a MPR of its smallest neighbor. They compared MPR-CDS with MPR algorithm described in [14]; the percentage of forwarding nodes in MPR algorithm was fewer than that in MPR-CDS by a minor amount.

In [11], the performance of a number of MPR algorithms was evaluated. The conclusions were: MPR based broadcasting schemes provide different features based on different MPR selection criteria that can be customized to obtain different broadcast performances as required. A gateway MPR (GMPR) algorithm was proposed in [26]. The simulation results showed that GMPR produces a smaller size CDS than the source-independent MPR in both sparse and dense networks. Also, an enhanced approach to the GMPR to further reduce the CDS size was presented in [27]. In [12] several extensions to generate smaller CDS were suggested. A comparison between the performance of MPRs and network coding (NC) algorithms was given in [28]. The comparison demonstrated that NC algorithm does not bring any benefits in terms of RET when compared to MPR algorithm.

## 3. WIRELESS NETWORK ENVIRONMENT

The wireless network environment can be categorized, according to the presence of noise or packet-loss, into two types of environments; these are [20]:

(1) A noiseless (error-free) environment, which represents an ideal network environment, in which it is assumed that all data transmitted by a source node is successfully and correctly received by a destination node. It is characterized by the following axioms or assumptions: the world is flat, all radios have equal range, and their transmission range is circular, communication link symmetry, perfect link, signal strength is a simple function of distance.

(2) A noisy (error-prone) environment, which represents a realistic network environment, in which the received signal will differ from the transmitted signal, due to various transmission impairments, such as: wireless signal attenuation, free space loss, thermal noise, atmospheric absorption, multipath effect, refraction.

All of these impairments are represented by a generic name, noise, and the environment is called noisy environment. For modeling and simulation purposes, the noisy environment can be described by introducing a probability function, which referred to as the probability of reception ($p_c$). It is defined as the probability that a wireless transmitted data is survived being lost and successfully delivered to a destination node despite the presence of all or any of the above impairments.

## 4. MPR ALGORITHMS

### 4.1. Concept of the MPR Algorithms

The idea behind MPR algorithms is to define, for each node in the network, a set of nodes called MPRs or simply relay nodes, these relay nodes are a subset of the one-hop neighbors of the node, which can establish communication paths with all two-hop neighbors. They are responsible for forwarding the broadcast message (e.g., RREQ packet) upon receiving it for the first time, while non relay nodes will not forward the message. The set of MPRs or relay nodes of a particular node ($x$) is referred to as *MPR*($x$) [25, 14].

The number of relay nodes in *MPR*($x$) is variable and it depends on the network topology, obviously it is less than or equal the number of one-hop neighbors. When the relay nodes are the same as the one-hop neighbors then this is pure flooding.

MPR algorithms require that each node knows the full list of its one-hop neighbor nodes ($N_1(x)$) and its two-hop neighbor nodes ($N_2(x)$). This information is collected via the periodic *HELLO* messages transmitted by mobile nodes. The *HELLO* messages contain the list of the one-hop nodes heard by the originator of the *HELLOs*. So that each node by collecting these *HELLO* messages can identify its one- and two-hop neighbor nodes, i.e., $N_1(x)$ and ($N_2(x)$).

Figure 1 illustrates how does an MPR algorithm works in a regular-geometry and noiseless environment. It shows that to diffuse a packet to the three-hops neighbors, a source node uniformly surrounded by 8, 16, and 24 one-, two-, three- hops neighbors, respectively, pure MPR algorithm needs 11 retransmissions as compared to 24 for pure flooding [14].

It can be clearly seen from Figure 1 that an MPR algorithm may reduce the number of redundant retransmissions at no cost of the network reachability. However, with high transmission errors, some of the forwarding nodes may not receive the packet due to a transmission error; this may result in a failure of delivery of the broadcast packet to all nodes in the network.

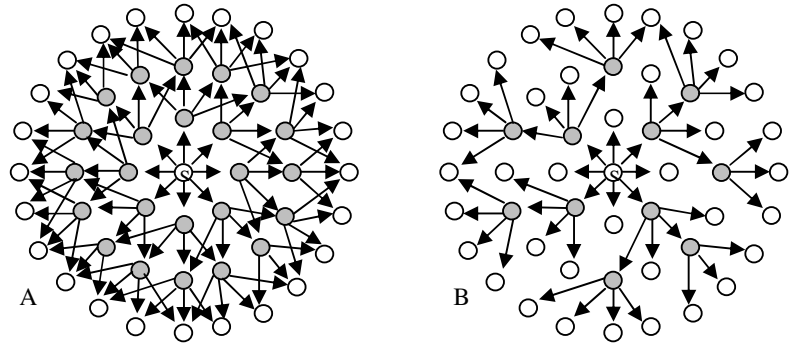

Figure 1. Diffusion of broadcast packet using: (A) pure flooding. (B) MPR flooding.

Figure 2A shows the use of MPR algorithms for flooding of a broadcast packet in a network that is characterized by a non-uniform node distribution and noiseless environment ($p_c$=1), while Figure 2B shows the flooding of a broadcast packet using MPR in a noisy environment. If using pure flooding, nodes F, G and H will have a chance to receive the packet from either node A, B, or C. While, using MPR nodes F, G, and H will have a chance to receive the packet from node B only. Thus, in a noisy environment, if the link between the source S and B is broken, then nodes F, G, and H will be isolated and no data can be delivered to them.

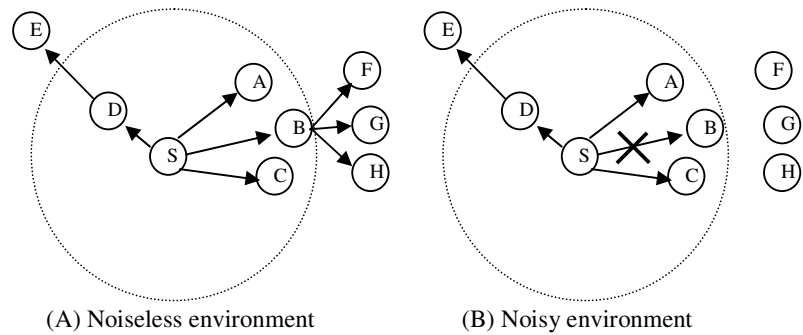

(A) Noiseless environment    (B) Noisy environment

Figure 2. Flooding using MPR algorithm in noiseless and noisy environments.

### 4.2. Costs of MPR Algorithms

In order to calculate the forwarding nodes, a certain number of procedures and information are required. These requirements form the cost of the MPR selection algorithm. Four costs of MPR algorithms described as follows [11]:

- **Time complexity:** is the time required to complete the forwarding nodes calculations. A heuristic that requires much time to run the calculation may be too complex to be deployed. Furthermore, when the network topology changes rapidly, the frequency of a forwarding node calculation also increases, and thus the time consumption of the calculation is huge for a complex heuristic. Hence, an efficient heuristic that consumes less time is essential for the MPR set generation.

- **Message complexity:** is the number of *HELLO* messages required for the calculation of the MPR set. For any MPR scheme, a number of *HELLO* messages need to be exchanged between nodes in advance. These *HELLO* messages contain the necessary information for a heuristic to implement the forwarding node set calculation. Algorithms in different groups or even in the same group may require a different number of *HELLO* messages. However, frequent information exchange will consume the limited bandwidth in MANETs and also accelerate the energy consumption of mobile nodes. Therefore, the number of *HELLO* messages exchanged, which is regarded as the message complexity, can significantly affect the performance of an MPR algorithm.

- **Information range:** is the hop level of neighboring nodes information (i.e. two-hops, three-hops, etc.) needed for the calculation of MPRs. Generally, the larger information range an algorithm requires, the more time and message exchange it will need depending on the algorithm. For example, an information range up to four hops may not be efficient for an MPR algorithm because messages need a long time to be transmitted to the source node and the information they carry may be outdated by then.

- **Source dependant:** in which a forwarding node need to know from which node the packet was received in order to determine whether or not to retransmit this packet. If an algorithm is not source dependant, a forwarding node will broadcast all messages that are received for the first time. This requirement increases the complexity of both the message sending and receiving process in an algorithm.

### 4.3. Pure MPR Algorithm

Pure MPR algorithm is designed to reduce the number of forwarding nodes and maintain the same network reachability, regardless of any other operation or optimization issues (e.g., QoS, reliability, power conservation, etc.). Therefore, the forwarding nodes selection heuristic is relatively simple. In addition, in pure MPR algorithm, using the neighbor-knowledge information obtained via periodic *HELLO* messages, each node can locally and independently calculate its own set of $N_1(x)$, $N_2(x)$ and $MPR(x)$ nodes. Figure 3 outlines a simple pure MPR algorithm [14].

---

**// The Heuristic Used for Selecting the MPRs in Pure MPR Algorithm**

For each node, after receiving *HELLO* messages from its neighbors, do:

(1) Construct its own set of $N_1(x)$ and $N_2(x)$ nodes
(2) Start with an empty MPR set $MPR(x)$
(3) First select those one-hop neighbor nodes in $N_1(x)$ as MPRs which are the only neighbor of some node in $N_2(x)$, and add these one-hop neighbor nodes to $MPR(x)$
(4) While there still exist some nodes in $N_2(x)$ which are not covered by $MPR(x)$:
   a. For each node in $N_1(x)$ which is not in $MPR(x)$, compute the number of nodes that it covers among the uncovered nodes in the set $N_2(x)$.
   b. Add the node of $N_1(x)$ to $MPR(x)$ for which this number is maximum.

---

Figure 3. The heuristic used for selecting the MPRs in pure MPR Algorithm [14].

To analyze the above pure MPR algorithm, first notice that step 3 permits to select some one-hop neighbor nodes as MPRs which must be in the $MPR(x)$ set. Otherwise the $MPR(x)$ will not cover all the two-hop neighbors. These nodes will be selected as MPRs in the process, sooner or later. Therefore, if step 3 is omitted, the MPRs set can still be calculated with success, i.e., it will cover all the two-hop neighbors. The presence of step 3 is for optimizing the MPR set. Those nodes which are necessary to cover the two-hop set $N_2(x)$ are all selected in the beginning, which helps to reduce the number of uncovered nodes of $N_2(x)$ at the start of the normal recursive procedure of step 4.

One drawback of pure MPR algorithm is that the selected forwarding set (i.e., $MPR(x)$) may not represent the optimum selection. This is because in step 4-a, there may be more than one nodes in $N_1(x)$ cover the same maximum number of nodes in $N_2(x)$. In step 4-b, one of nodes that covers this maximum number is selected, for example, by considering the node's ID, which not enough to ensure the optimum selection.

## 5. THE PROPOSED OPTIMAL MPR (OMPR) ALGORITHM

### 5.1. A New MPRs Selection Heuristic

In this section, we describe a new heuristic for selecting the optimal $MPR(x)$ set, which is the set that has the minimum number of forwarding nodes. An optimal $MPR(x)$ set for a node is defined as a subset of the one-hop neighbors, which covers the two-hop neighbors of that node, and it has the minimum number of nodes among all other sets that cover the two-hop neighbors of the node.

The new heuristic can be summarized as follows: in step 4-b in Figure 3, if more than one nodes in $N_1(x)$ can cover the same maximum number of nodes in $N_2(x)$, then for each node in $N_1(x)$, we generate a new $MPR(x)$ set and each generated set will be processed independently until all nodes in $N_2(x)$ are covered by the generated $MPR(x)$ set. Thus we will end up with a number of valid $MPR(x)$ sets. The $MPR(x)$ set selected is the one with minimum number of nodes. Figure 4 outlines the heuristic used for selecting the optimal $MPR(x)$ set.

---

**// The Heuristic Used for Selecting the Optimal $MPR(x)$ set**
For each node, after receiving *HELLO* messages from its neighbors, do:
  (1) Construct its own set of $N_1(x)$ and $N_2(x)$ nodes
  (2) Start with an empty MPR set $MPR(x)$
  (3) First select those one-hop neighbor nodes in $N_1(x)$ as MPRs which are the only neighbor of some node in $N_2(x)$, and add these one-hop neighbor nodes to the MPR set $MPR(x)$
  (4) Set $M=1$, where $M$ is the number of $MPR(x)$ sets
  (5) Set $m=1$, where $m$ is a counter for the number of $MPR(x)$ sets
  (6) While ($m \leq M$)
    a. While there still exist some nodes in $N_2(x)$ which are not covered by $MPR(x)_m$:
      i. For each node in $N_1(x)$ which is not in $MPR(x)_m$, compute the covered nodes
      ii. Compute the maximum number of nodes among the uncovered nodes in the set $N_2(x)$ that is covered one or more nodes in the set $N_1(x)$
      iii. Compute the number of nodes ($K$) which cover this maximum
      iv. If ($K=1$)
          Add the node of $N_1(x)$ to $MPR(x)_m$
        Else ($K>1$)
          Compute the number of new generated set ($A$) as: $A=K-1$
          Select a node in $N_1(x)$ that covers this maximum
          Add the node of $N_1(x)$ to $MPR(x)_m$
          For ($i=1$ to $A$) (Each other node that covers this maximum)
            Generates new set as follows:
            Select a node in $N_1(x)$ that covers this maximum
            Add the node of $N_1(x)$ to $MPR(x)_{M+i}$
          Compute the total number of sets ($M$) as: $M=M+K-1$
        End If
    b. $m=m+1$
  (7) For all sets $MPR(x)_1$ to $MPR(x)_M$ select the MPR set with the least number of nodes as $MPR(x)$.

Figure 4. The heuristic used for selecting the optimal MPRs in OMPR algorithm.

### 5.2. The Proposed OMPR Algorithm

The proposed MPR algorithm uses the heuristic described in Section 5.1 to select the optimum $MPR(x)$ set; therefore it is referred to as optimal MPR (OMPR) algorithm. The OMPR algorithm maximizes the performance of the network as it reduces the number of retransmission, which consequently reduces the bandwidth and power consumption, contention, and collisions at the receiver.

The route discovery OMPR algorithm can be described as follows: Each node within the network calculates its $MPR(x)$ set using the heuristic described in Section 5.1. Thus, each time a node receives a RREQ packet from its neighbors, it checks to see if it is used as a relay node for this neighbor. If so, then it forwards the packet, otherwise, it just discards it.

Figure 5 outlines the implementation of the proposed OMPR algorithm in MANETs to calculate two main performance measures, namely: (1) Number of retransmission (RET), which is defined as the average number of nodes that retransmit the RREQ normalized to the total number of nodes within the network (*n*); and (2) Reachability (RCH), which represents the average number of reachable nodes by any node within the network normalized to *n*, or it also can be defined as the probability that a RREQ packet initiated by any node (source node) will be successfully delivered to any other node (destination node) within the network.

Since each node needs to know its one-hop neighbor, in Figure 6 we present an algorithm for calculating the one-hop neighbor for each node in a noisy MANETs. This is done by presenting the noise level in terms of pc. So that, for each node to discover its one-hop neighbor, it must be sure that the node is within its transmission range and the $p_c$ is equal to less than a certain random number $\xi$. However, the same algorithm can be used for noiseless environment by setting $p_c$ to unity.

---

**// Calculation of RET and RCH Using the OMPR Route Discovery Algorithm**
(1) Generate nodes
(2) Calculate pause time ($\tau$=0.75*R/u) // R is the transmission radius and u is the average node speed
(3) Calculate number of mobility loops (M=$T_{sim}$/$\tau$) // Where $T_{sim}$ is the simulation time
(4) For (*m*=1 to M) Do
    a. For each node *x* (*x*=1, 2, …, *n*) // Where *n* is the number nodes within the network
        Calculate node one-hop neighbors $N_1(x)$ using the algorithm in Figure 6
    b. For each node *x* (*x*=1, 2, …, *n*)
        Use the heuristic described in Section 5.1 to select the optimal MPR(*x*) set
    c. For each node *x* (*x*=1, 2, …, *n*)
        Initiate RREQ
        Identify all nodes which are reachable by node *x*
        Construct the spanning tree starting with node *x*
        Use the MPR(*x*) set
          Calculate the total number of nodes which receive the RREQ packet (*Rx*)
          Calculate the total number of nodes which retransmit the RREQ packet (*Tx*)
        Accumulate *Rx*/(*n*-1) into *accRx* // The source node is excluded
        Accumulate *Tx*/(*n*-1) into *accTx* // The source node is excluded
    d. Calculate average value *accRx/n* and accumulate the result into *avgRx*
    e. Calculate average value *accTx/n* and accumulate the result into *avgTx*
    f. Update nodes location
(5) Calculate average value *avgRx/m* // Represents reachability (RCH)
(6) Calculate average value *avgTx/m* // Represents number of retransmission normalized to *n* (RET)

Figure 5. An algorithm for calculating RET and RCH using the OMPR algorithm.

---

**// Calculation of One-Hop Neighbor in a Noisy MANET.**
For a source node (*i*) Do:
    For each node within the network (except the source node *i*) Do:
        Calculate the distance (*d*) between two nodes (*i* and *j*) using $d = \sqrt{(x_i - x_j)^2 + (y_i - y_j)^2}$
        If (*d*≤R) Then                       // Where R is the source node transmission radius
            Generate a random number ($\xi$) (0≤$\xi$<1)
            If ($\xi$≤$p_c$) Then            // Where $p_c$ is the probability of reception
                Node is successfully delivered to node *j*
            Else ($\xi$>$p_c$)
                Node is failed to delivered to node *j*
            End If
        End If

Figure 6. Calculation of one-hop neighbor in a noisy MANET.

### 5.3. Features of the OMPR Algorithm

The main features of the OMPR algorithm can be summarized as follows:
- It is a neighbor-knowledge algorithm, in which each node needs to know the full list of its one-hop neighbors, and should pass this information back to all of them.
- Each node calculates its MPR set locally. Therefore, it is a source-dependent process.
- It generates an optimal MPR set for each node which has the minimum number of nodes among all other sets that cover the two-hop neighbors for that node.
- The number of relay nodes for each node depends on the network topology and it is highly affected by the node mobility.
- The performance of the OMPR may be extremely affected by the presence of noise. This is due to the fact that if the link between a source node and a relay node (a node in $N_1(x)$) is broken, then all nodes in $N_2(x)$ that are attained through this relay node are disconnected.

## 6. SIMULATIONS AND RESULTS

The network simulator used in this work is MANSim [19, 20]. It is a MANET simulator especially developed to simulate and evaluate the performance of a number of flooding optimization algorithms for MANETs. It is written in C++ language, and it consists of four major modules, these are: network, mobility, computational, and algorithm modules.

In order to evaluate and compare the performance of the proposed OMPR algorithm in noisy MANETs, a number of simulations were performed using MANSim. These simulations investigate the variation of RET and RCH with $p_c$. The simulation results are compared against those obtained by using a number of other flooding optimization algorithms such as: pure flooding, probabilistic flooding with fixed retransmission probabilities ($p_t$=0.8), probabilistic flooding with dynamic $p_t$, LAR-1, and LAR-1P with $p_t$=0.8. The input parameters for these simulations are listed in Table 1. The simulations results are plotted in Figures 7 and 8.

Table 1. Input parameters.

| Parameters | Values |
|---|---|
| Geometrical model | Random node distribution |
| Network area | 1000x1000 m |
| Number of nodes ($n$) | 100 nodes. |
| Transmission radius ($R$) | 200 m |
| Average node speed ($u$) | 5 m/sec |
| Probability of reception ($p_c$) | From 0.5 to 1.0 in step of 0.1 |
| Simulation time ($T_{sim}$) | 300 sec |
| Pause time ($\tau$) | $\tau$=0.75*($R/u$)=30 sec |

The main points that are concluded from these simulations can be summarized as follows:

- The probabilistic approach always achieves the highest possible RCH, but at the same time it introduces a low reduction in RET when it is compared with the other techniques.
- The LAR-1 and LAR-1P algorithms presents the highest reduction in RET but at the same time they provide the lowest RCH.
- The OMPR algorithm presents a moderate reduction in RET, when it is compared with probabilistic (fixed and dynamic $p_t$), LAR-1, and LAR-1P. It performs better than probabilistic and less than LAR-1 and LAR-1P for various values of $p_c$. The RCH it achieves is higher than that of LAR-1 and LAR-1P algorithms.
- The RCH of the OMPR algorithm is highly affected and ruined due presence of noise as shown in Figure 8.

Since the main objective of using flooding optimization during route discovery is to achieve a cost-effective RCH, which means a highest possible reachability at a reasonable number of retransmission. The obtained results demonstrated that the OMPR algorithm provides an excellent performance as it can achieve an excellent cost-effective reachability, for various network noise levels, as compared to other route discovery algorithms.

Figure 8 shows that the probabilistic and OMPR algorithms provide almost a comparative performance in noiseless and low-noise environments ($p_c>0.8$). But, in terms of network reachability, the probabilistic approach overwhelmed the performance of the OMPR algorithm in noisy environment. For example, for mobile nodes with $u$=5 m/sec and $p_c$=0.5, the OMPR algorithm achieves a reachability of only 34.6%, while for the same environment, the probabilistic approach achieves over 85%. But, the probabilistic approach achieves this high network reachability at a high cost of RET (≈68%) compared with RET=10.3% for OMPR.

Figures 7 and 8 demonstrate that the OMPR algorithm provides an excellent network RCH in noisy environment, when compared with LAR-1 and LAR-1P. For example, when $p_c$=0.8, OMPR achieves a RCH of 86.6% compared with 69.2% and 53.4% for LAR-1 and LAR-1P, respectively. However, this is achieved at a cost of 28.8% RET compared with 13.8% and 8.9%, for LAR-1 and LAR-1P, respectively. The results also demonstrate that the OMPR algorithm is very sensitive to the variation in noise-level. Pure flooding is the least affected algorithm, then probabilistic algorithm, followed by the LAR-1 and LAR-1P algorithms.

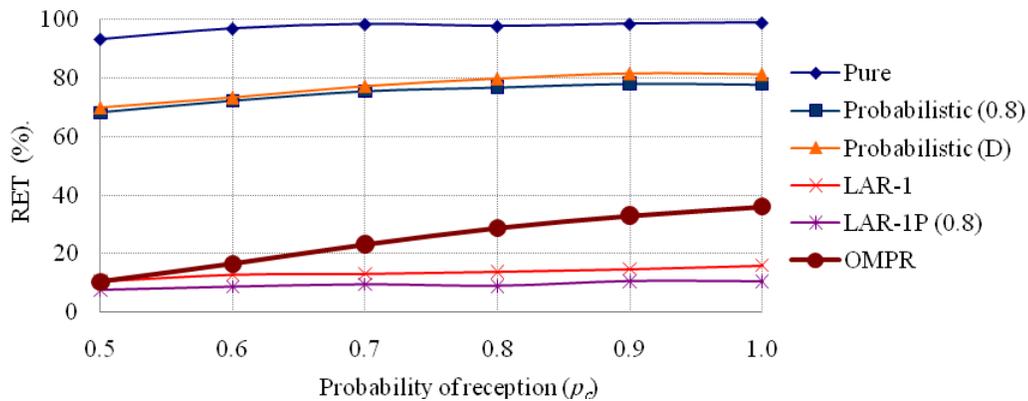

Figure 7. Variation of RET with $p_c$ for various algorithms.

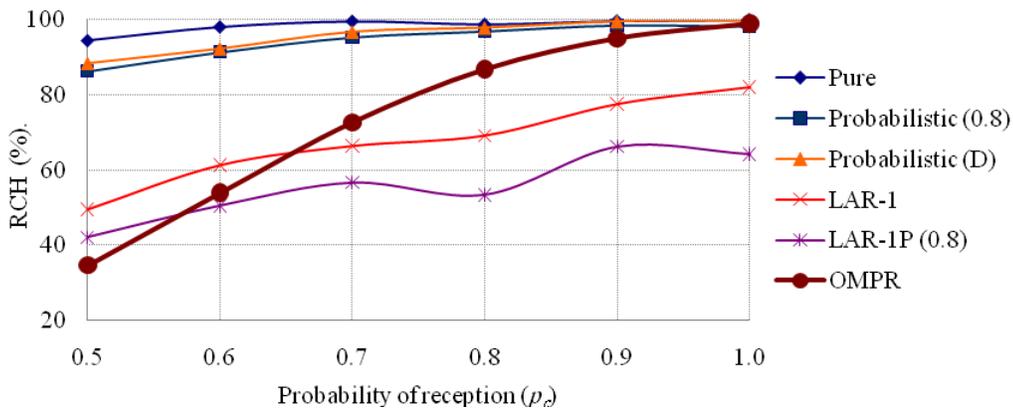

Figure 8. Variation of RCH with $p_c$ for various algorithms.

## 7. CONCLUSIONS

The main conclusion of this work is that the OMPR algorithm demonstrated an excellent cost-effective performance as compared to other route discovery algorithms, such as: pure flooding, probabilistic flooding, LAR-1 and LAR-1P algorithms In particular, OMPR provides a satisfactory RCH as compared to pure and probabilistic algorithms in high mobility noiseless and low-noise level MANETs environment ($p_c \geq 0.8$), and always higher than LAR-1 and LAR-1P algorithms. Also, OMPR significantly reduces RET while maintaining an appropriate RCH in various MANETs environments. The simulation results demonstrate that the performance of the OMPR algorithm is equivalent to the performance of pure flooding with minimum cost, when $p_c > 0.8$. The main drawback of the OMPR is its high sensitivity to noise-level as it yields the highest average rate of change in reachability in comparison with other algorithms.

Our main recommendations for future work are to consider other factors during the selection of MPR set, such as: nodes residue energy level, nodes reliability, nodes security measures, etc. Also, in order to enhance the performance of the OMPR algorithm in noisy MANETs, we suggest modifying the heuristic to add supporting nodes to the MPR set to maintain high reachability in such environment. Furthermore, it is important to perform further investigation to the performance, for example, investigate the performance of the algorithm under realistic mobility models and variable node densities and nodes radio transmission range.

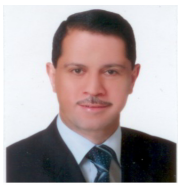
***Dr. Hussein Al-Bahadili*** (*Hbahadili@aabfs.org*) is an associate professor at the Arab Academy for Banking & Financial Sciences (AABFS). He earned his M.Sc and PhD from the University of London (Queen Mary College) in 1988 and 1991, respectively. He received his B.Sc in Engineering from the University of Baghdad in 1986. He is a visiting researcher at the Centre of Wireless Networks and Communications (WNCC) at the School of Engineering, University of Brunel (UK). He has published many papers in different fields of science and engineering in numerous leading scholarly and practitioner journals, and presented at leading world-level scholarly conferences. He recently published two Chapters in two prestigious books in IT and Simulations. His research interests include computer networks design and architecture, routing protocols optimizations, parallel and distributed computing, cryptography and network security, data compression, software and web engineering.

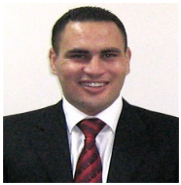
***Mr. Rami Jaradat*** (*Rami@hu.edu.jo*) is an IT Lab Demonstrator at the Hashemite University, Jordan. He received his B.Sc degree in Electrical and Computer Engineering from Jordan University of Science and Technology, Jordan in 2002, and his M.Sc degree in Computer Science from Amman Arab University for Graduate Studies, Jordan in 2009. His current research interests are in developing efficient dynamic routing protocols for mobile ad hoc networks, wireless networks management and security, and ad hoc networks modeling and simulation.